\documentclass{llncs}

\usepackage{amsmath,amssymb,amsfonts}
\usepackage{paralist}
\usepackage{alltt}
\usepackage{wrapfig}

\usepackage{caption}
\usepackage{subcaption}
\usepackage{subfig}
\usepackage{wrapfig}

\usepackage{algorithm}
\usepackage[noend]{algorithmic}

\usepackage{tikz}
\usetikzlibrary{matrix}
\usetikzlibrary{fit}
\usetikzlibrary{automata}
\usetikzlibrary{shapes}
\usetikzlibrary{arrows,positioning}
\usepackage{pgfplots} 

\newcommand\comment[1]{}
\renewcommand\paragraph[1]{\noindent{\bf #1}}

\newcommand\arsays[1]{{\bf AR: #1}}

\newcommand{\esolver}{{\sc eSolver}}
\newcommand{\PUFFIN}{{\sc Puffin}}
\newcommand{\RAZORBILL}{{\sc Razorbill}}
\newcommand{\AUK}{{\sc Auk}}

\newcommand{\ProgSpace}{\mathcal{P}}
\newcommand{\expr}{\mathtt{expr}}
\newcommand{\Valuation}{\mathcal{V}}
\newcommand{\concreteConstraint}{\rho}
\newcommand{\InpSpace}{\mathcal{I}}
\newcommand{\OutputVariable}{o}
\newcommand{\Prog}{\mathtt{Prog}}
\newcommand{\SubProg}{\mathtt{SubProg}}
\newcommand{\inp}{\mathtt{inp}}

\newcommand{\True}{\mathtt{true}}
\newcommand{\None}{\mathtt{None}}

\newcommand{\constraints}{\varphi}
\newcommand{\UnifConstraints}{\psi}
\newcommand{\BackConstraints}{\beta}
\newcommand{\splitInpSpace}{\mathit{splitInpSpace}}

\newcommand{\Spec}{\mathit{Spec}}
\newcommand{\calI}{\mathcal{I}}
\newcommand{\calJ}{\mathcal{J}}
\newcommand{\calK}{\mathcal{K}}
\newcommand{\calP}{\mathcal{P}}

\newcommand{\TimeOut}{\mathtt{timeOut}}
\newcommand{\Not}{\mathtt{Not}}

\newif\iffull

\fulltrue

\begin{document}
\title{Synthesis through Unification
  \thanks{This research was supported in part by the NSF under award CCF
    1421752 and the Expeditions award CCF 1138996, by DARPA under
    agreement FA8750-14-2-0263, by the Simons
    Foundation, and by a gift from the Intel Corporation.}
} 
\author{
  Rajeev Alur\inst{1} \and
  Pavol {\v C}ern{\'y}\inst{2} \and
  Arjun Radhakrishna\inst{1}
}
\institute{
  University of Pennsylvania \and
  University of Colorado Boulder 
}
\maketitle

\begin{abstract}
Given a specification and a set of candidate programs (program space),
the program synthesis problem is to find a candidate program that
satisfies the specification.
We present the synthesis through unification (STUN) approach, which is
an extension of the  
counter-example guided inductive synthesis (CEGIS) 
approach. In CEGIS, the synthesizer maintains a subset $S$ of inputs 
and a candidate program $\Prog$ that is correct for $S$. The synthesizer
repeatedly checks if there exists a  
counterexample input $c$ such that the execution of $\Prog$ is incorrect
on $c$.  If so, the synthesizer enlarges $S$ to include $c$, and picks
a program from the program space that is correct for the new set $S$. 

The STUN approach extends CEGIS with the
idea that given a  program $\Prog$ that is correct for a subset of
inputs, the synthesizer can try to find a program $\Prog'$ that is
correct for the rest of the inputs. If $\Prog$ and $\Prog'$ can be {\em
unified} into a program in the program space, then a solution has
been found. 
We present a generic synthesis procedure based on the STUN approach
and specialize it for three different domains by providing the
appropriate unification operators. 
We implemented these specializations in prototype tools, and we show
that our tools often performs significantly better on standard
benchmarks than a tool based on a pure CEGIS approach.
\end{abstract}

\vspace{-5ex}
\section{Introduction}
\label{sec:intro}
\vspace{-1ex}

The task of program synthesis is to construct a program
that satisfies a given declarative specification.
The computer-augmented programming~\cite{asplos06,sygus} approach allows
the programmers to express their intent in different ways, for 
instance by providing a partial program, or by defining the space of
candidate programs, or by providing positive and negative examples and
scenarios.
This approach to synthesis is becoming steadily more popular
and successful~\cite{survey}. 

We propose a novel algorithmic approach for the following problem:
given a specification, a set of candidate programs (a program space),
and a set of all possible inputs (an input space), find a candidate
program that satisfies the specification on all inputs from the input
space. 
The basic idea of our approach is simple: if we have a candidate
program that is correct only on a part of the input space, we can
attempt to find a program that works on the rest of the input space,
and then unify the two programs. The unification operator must ensure
that the resulting program is in the program space. 

The program space is syntactically restricted
to a set which can be specified using a typed grammar.
If this grammar 
contains {\tt if} statements, and its expression language is
expressive enough, then a simple unification operator exists. A
program $\Prog$ for inputs that satisfy an expression
$C$, and a program  
$\Prog'$ that works on the rest of the inputs can be unified into 
$\mathsf{if}~(C)~\mathsf{then}~\Prog~\mathsf{else}~\Prog'$.
Even when $\mathsf{if}$ statements are not available, different
unification operators may exist.
These unification operators may be preferable to
unification through $\mathsf{if}$ statements due to efficiency reasons.
However, such unification operators may not be
complete --- it might not be possible to unify two given programs. 
We present an approach that deals with such cases with appropriate
backtracking. 

Our approach, which we dub STUN, works as follows: its first step is
to choose a program $\Prog$ that works for a subset $\InpSpace_G$ of the
input space.
This step can be performed by any existing method, for instance
by multiple rounds of the CEGIS loop~\cite{SL13}. 
The STUN procedure then makes a recursive call to itself 
to attempt to synthesize a program $\Prog'$ for inputs on which $\Prog$ is
incorrect. 
An additional parameter is passed to the recursive call --- 
unification constraints that ensure that the program $\Prog'$ obtained from the
recursive call is unifiable with $\Prog$.
If the recursive call succeeds, programs $\Prog$ and $\Prog'$ can be
unified, and the solution to the original problem was found.
If the recursive call fails, then we need to backtrack, and choose
another candidate for program $\Prog$. In this case, we also use a form of
conflict-driven learning. 

\noindent{\bf Problem domains.}
We instantiate the STUN approach to three different problem domains:
bit-vector expressions, separable specifications for conditional
linear arithmetic expressions, and non-separable specifications for
conditional linear arithmetic expressions. 
In each domain, we provide a suitable unification operator,
and we resolve the nondeterministic choices in
the STUN algorithm. 

We first consider the domain of bit-vector expressions.
Here, the challenge is the absence 
of {\tt if}-conditionals, which makes the unification operator harder
to define. 
We represent bit-vector programs as
$(\expr,\concreteConstraint)$, where $\expr$ is a bit-vector expression
over input variables and additional auxiliary variables,
and $\concreteConstraint$ is a constraint over the auxiliary variables.
Two such pairs $(\expr_1,\concreteConstraint_1)$ and
$(\expr_2,\concreteConstraint_2)$ can be unified if there exists a way
to substitute the auxiliary variables in $\expr_1$ and
$\expr_2$ to make the expressions equal, and the substitution
satisfies the conjunction of 
$\concreteConstraint_1$ and $\concreteConstraint_2$. 
A solver based on such a unification operator has comparable
performance on standard benchmarks~\cite{syguscomp} as existing solvers. 

For the second and third domain we consider, the program space is the
set of conditional linear-arithmetic expressions (CLEs) over
rationals. The difference between the two domains is in the form of
specifications. Separable specifications are those where the 
specification only relates an input and its corresponding output.
In contrast, the non-separable specifications can place constraints
over outputs that correspond to different inputs. For instance,
$x>0 \implies f(x+2) = f(x) + 7$ is a non-separable specification, as it
relates outputs for multiple inputs.

The second domain of separable specifications and CLEs over rationals
is an ideal example for STUN, as the unification operator is easy to
implement using conditions of CLEs. 
We obtain an efficient implementation where partial solutions are
obtained by generalization of input-output examples,
and such partial solutions are then unified.
Our implementation of this procedure is order-of-magnitude faster on
standard benchmarks than the existing solvers.

The third domain of non-separable specifications for CLEs requires 
solving constraints for which finding a solution might need an unbounded
number of 
unification steps before convergence. 
We therefore implement a widening version of the
unification operator, further demonstrating the generality of the
STUN approach. 
Our implementation of this procedure performs on par with existing
solvers on standard benchmarks.

\noindent{\bf Comparing CEGIS and STUN.}
The key conceptual difference between existing synthesis methods
(CEGIS) and our STUN approach is as 
follows: CEGIS gradually collects a set of input-output
examples (by querying the specification), and then finds a solution
that matches all the examples. The STUN approach also collects 
input-output examples by querying the specification, but it finds a
(general) solution for each of them 
separately, and then unifies the solutions. The STUN method has an
advantage if solutions for different parts of the input space are
different. In other words, CEGIS first combines subproblems, and then
solves, while STUN first solves, and then combines solutions. 
The reason is that such solutions can be in many cases
easily unifiable (if for instance the program space has {\tt
  if} conditionals), but finding the whole solution at once for
examples from the different parts of input space (as CEGIS
requires) is difficult. 

\noindent{\bf Summary.}
The main contributions of this work are two-fold.
First, we propose a new approach to program synthesis based on
unification of programs, and we develop a generic 
synthesis procedure using this approach. 
Second, we instantiate the STUN synthesis procedure to the domains of
  bit-vector expressions, and conditional linear expressions with
  separable and non-separable specifications. We show that in all
  cases, our solver has comparable performance to existing solvers,
  and in some cases (conditional linear-arithmetic expressions with
  separable specifications), the
  performance on standard benchmarks is several
  orders of magnitude better. This demonstrates the potential of the
  STUN approach. 

\comment{
CAV2015: 
\begin{itemize}
 \item 15 pages, not counting references
 \item Abstract due: 31th January, 13:00 Vienna
 \item Paper due: 7th February, 13:00 Vienna
 \item Artifact evaluation {\bf after} acceptance
\end{itemize}

In the popular counter-example guided inductive synthesis (CEGIS)
approach, the synthesizer repeatedly picks a candidate
program and checks if there exists an input such that the candidate
program's behaves incorrectly on the input.
If there is no such input, the candidate program is returned.
Otherwise, the synthesizer eliminates from the program space all the
programs that behave incorrectly on that particular input.

While CEGIS has been very successful in different domains, the method
has some drawbacks. 
In particular, it does not consider the structure of programs, but
instead treats them as monolithic objects, i.e., a program is boiled
down a single bit of information, i.e, whether or not it is correct on
all inputs.

Further, we also show that several other synthesis algorithms in the
literature are also specializations of the STUN approach. (say
which)

* CEGIS and SYGUS
* abstraction-guided synthesis (the concurrency idea)
* our work  --concurrency - small atomic sections fixes  a schedule...
* skolem functions - FMCAD 2014 - armin biere - 
* version-space algebra - combines things for an input

** abduction 
}

\vspace{-2ex}
\section{Overview}
\label{sec:illustrative_example}
\vspace{-1ex}

In this section, we first present a simplified view of synthesis by
unification (the UNIF loop), which works under very strong
assumptions. We then describe what extensions are needed, and motivate
our STUN approach.

\paragraph{UNIF loop.}
Let us fix a specification $\Spec$, a {\em program space} $\calP$ (a set
of candidate programs), and an {\em input space} $\calI$.
The program synthesis problem is to find a program in $\calP$ that
satisfies the specification for all inputs in $\calI$.

A classical approach to synthesis is the counterexample-guided
inductive synthesis (CEGIS) loop. We choose the following presentation
for CEGIS in order to contrast it with UNIF.   
In CEGIS (depicted in Figure~\ref{fig:cegis}), the synthesizer maintains
a subset $\calJ \subseteq \calI$ of inputs and a candidate program
$\Prog \in \calP$ that is correct for
$\calJ$. 
If $\calJ = \calI$, i.e., if $\Prog$ is correct for all inputs
in $\calI$, the CEGIS loop terminates and returns $\Prog$. 
If there is an input on which $\Prog$ is incorrect, the first
step is to find such an input $c$.
The second step is to find a program that is correct
for both $c$ and all the inputs in $\calJ$. 
In Figure~\ref{fig:cegis}, this is done in the call to {\tt
  syntFitAll}). This process is then repeated until $\calJ$ is equal to
$\calI$. 

The unification approach to synthesis is based on a simple
observation: if we have a program $\Prog$ that is correct for a subset $\calJ$
of inputs (as in CEGIS), the synthesizer can try to find a program
$\Prog'$ that is correct for some of the inputs in $\calI \setminus \calJ$,
and then attempt to unify $\Prog$ and $\Prog'$ into a program in the program
space $\calP$. We call the latter option the UNIF loop. It is depicted
in Figure~\ref{fig:unif}. In more detail,
the UNIF loop works as follows. We first call {\tt syntFitSome} in
order to synthesize a program $\Prog'$ that works for some inputs in
$\calI$ but not in $\calJ$. Let $\calJ'$ be the set of those inputs in $\calI
\setminus \calJ$ for which $\Prog'$ satisfies $\Spec$. 

Next, we consider two programs $\calJ \cdot \Prog$ and $\calJ' \cdot
\Prog$, where the notation $\calJ \cdot \Prog$ denotes a program that on
inputs in $\calJ$ behaves as $\Prog$, and on 
other inputs its behavior is undefined. We need to unify the two
programs to produce a program (in the program space
$\calP$) which is defined on $\calJ \cup \calJ'$. The unification operator 
denoted by $\oplus$, and the unified program is obtained as
$\calJ \cdot \Prog \oplus \calJ' \cdot \Prog$. If the program space is
closed under {\tt if} conditionals, and if $\Prog$ and $\Prog'$ are in $\calP$,
then the unification is easy. We obtain {\tt if} $\calJ$ {\tt then} $\Prog$
{\tt else if} $\calJ'$ {\tt then} $\Prog'$ {\tt else} $\bot$. Note
that we abuse notation here: the symbols $\calJ$ and $\calJ'$, when used in
programs, denote expressions that define the corresponding input
spaces. 

\begin{figure}[t]%
  \vspace{-3ex}
\begin{minipage}{0.42\textwidth}%
\begin{tikzpicture}[
font=\scriptsize, node distance=4mm, >=latex',
block/.style = {draw, rectangle, 
              align=center}, 
rblock/.style = {draw, rectangle, rounded corners=0.5em, align=center},
tblock/.style = {draw, trapezium, 
                 trapezium left angle=75, trapezium right angle=105,
                 align=center}, 
dblock/.style = {draw, diamond, align=center, inner sep=1pt}
                      ]
    \node [tblock]           (candidate) 
          {$(\Prog,\calJ)$ \\ such that \\ $\forall j: j \in \calJ \Leftrightarrow
            (\Prog(j) \models \Spec)$}; 
    \node [dblock, below right= of candidate, xshift=12mm]  (done?)   {$\calJ \stackrel{?}{=} \calI$};
    \node [block, below= of done?, xshift=2mm]  (return)   {{\tt return} $\Prog$};
    \node [block, left= of done?]     (ctrex)  {$c \gets$ {\tt
        ctrex}($\Prog,\Spec$)}; 
    \node [block, below=of ctrex]     (fitall)  {$(\calJ,\Prog) \gets $ \\ 
                                   {\tt syntFitAll}($\calJ \cup \{c\},\Spec$)}; 
   

    \path[draw,->] (candidate)   edge (done?)
                   (done?)    edge node[right] {yes} (return)
                   (done?)   edge node[above] {no} (ctrex)
                   (ctrex)   edge (fitall)
                    ;

    \path[draw, rounded corners, ->] (fitall.west) 
          -- +(-0.2cm, 0cm)
          -- +(-0.2cm,13mm)
	-- (candidate.south west);

\end{tikzpicture}
\caption{CEGIS loop for input space $\calI$ and specification $\Spec$}
\label{fig:cegis}
\end{minipage}
\qquad
\begin{minipage}{0.45\textwidth}%
\begin{tikzpicture}[
font=\scriptsize, node distance=4mm, >=latex',
block/.style = {draw, rectangle, 
              align=center}, 
rblock/.style = {draw, rectangle, rounded corners=0.5em, align=center},
tblock/.style = {draw, trapezium, 
                 trapezium left angle=75, trapezium right angle=105,
                 align=center}, 
dblock/.style = {draw, diamond, align=center, inner sep=1pt}
                      ]
    \node [tblock]           (candidate) 
          {$(\Prog,\calJ)$ \\ such that \\ $\forall j: j \in \calJ \Leftrightarrow
            (\Prog(j) \models \Spec)$}; 
    \node [dblock, below left= of candidate, yshift=-1mm]  (done?)   {$\calJ \stackrel{?}{=} \calI$};
    \node [block, below= of done?]  (return)   {{\tt return} $\Prog$};
    \node [block, right= of done?, xshift=3mm] (fitsome)  
                               {$(\calJ',\Prog') \gets$ \\
                            {\tt syntFitSome}($\calI \setminus \calJ, \Spec$)}; 
    \node [block, below=of fitsome]  (unif)  {$\calJ \gets \calJ \cup \calJ'$ \\ 
                                       $\Prog \gets (\calJ \cdot \Prog) \oplus
                                        (\calJ' \cdot \Prog')$}; 
    

    \path[draw,->] (candidate)   edge (done?)
                   (done?)    edge node[right] {yes} (return)
                   (done?)   edge node[above] {no} (fitsome)
                   (fitsome)   edge (unif)
                    ;

    \path[draw, rounded corners, ->] (unif.east) 
          -- +(2mm,0cm)
          -- +(2mm,15mm)
	-- (candidate.east);


\end{tikzpicture}
\caption{UNIF loop for input space $\calI$ and specification $\Spec$}
\label{fig:unif}%
\end{minipage}%
\end{figure}%

\begin{example}
\vspace{-1ex}
Consider the following specification for the function $\max$. 
\begin{center}
\vspace{-2ex}
$\Spec = f(x,y) \geq x \wedge f(x,y) \geq y \wedge (f(x,y) = x \vee
f(x,y) = y)$
\vspace{-2ex}
\end{center}

\noindent The input space $\calI$ is the set of all pairs of integers. 
The program space $\calP$ is the set of all programs in a simple
if-language with linear-arithmetic expressions. 

We demonstrate the UNIF loop (Figure~\ref{fig:unif}) on this example. 
We start with an empty program $\bot$. The program works for no
inputs (i.e., the input space is $\emptyset$), so we start with the
pair $(\bot,\emptyset)$ at the top of Figure~\ref{fig:unif}. As $\emptyset 
\neq \calI$, we go to the right-hand side of Figure~\ref{fig:unif},
and call the procedure {\tt syntFitSome}. 

We now describe the procedure {\tt syntFitSome($\calK,\Spec$)} for the linear
arithmetic domain. It takes two parameters: a set of inputs $\calK$, and a
specification $\Spec$, and returns a pair $(\calJ',\Prog')$ consisting of a
set $\emptyset \neq \calJ' \subseteq \calK$ and 
a program $\Prog'$ which is correct on $\calJ'$. 
We pick an input-output example from the input space $\calK$.  
This can be done by using a satisfiability solver to obtain a model of
$\Spec$. 
Let us assume that the specification is in CNF. An input-output example
satisfies at least one atom in each clause. Let us pick those atoms. 
For instance, for the example $(2,3) \rightarrow 3$, we get the
following conjunction $G$ of atoms: 
  $G \equiv f(x,y) \geq x \wedge f(x,y) \geq y \wedge f(x,y) = y$. 
We now generate a solution for the input-output example and $G$. For linear
arithmetic, we could ``solve for $f(x,y)$'', i.e. replace $f(x,y)$ by
$t$ and solve for $t$. Let us assume that the solution $\Prog_0$ that we
obtain is a function that on any input $(x,y)$ returns $y$. 
We then plug the 
solution $\Prog_0$ to $G$, and simplify the resulting formula in order to
obtain $G_0$, where $G_0$ is $y \geq x$. 
$G_0$ defines the set of inputs for which the solution is correct. 
We have thus effectively obtain the pair $(G_0,\Prog_0)$ that the function
returns (this effectively represents the program 
  {\tt if} $y \geq x$ {\tt then} $y$ else $\bot$).

In the second iteration, we now call the function {\tt
  syntFitSome($\calK,\Spec$)} with the parameter $\calK = \neg G_0$. 
We now ask for an input-output example where the input satisfies $\neg
G_0$.   
Let us say we obtain $(5,4)$, with output $5$. 
By a similar process as above, we obtain a program $\Prog_1$
that for all inputs $(x,y)$ returns $x$, and works for all input that
satisfy $G_1 \equiv x \geq y$.

The next step of the STUN loop asks us to perform the unification 
$(G_0 \cdot \Prog_0) \oplus (G_1 \cdot \Prog_1)$. Given that we have {\tt if} conditionals
in the language, this step is simple.  
We unify the two programs to obtain: 
{\sf if} $y \geq x$ {\sf then} $y$ {\sf else} $x$.
\vspace{-1ex}
\end{example}

\paragraph{From the UNIF loop to STUN}
The main assumption that the UNIF loop makes is that the unification
operator $\oplus$ always succeeds. We already mentioned that 
this is the case when the program space is closed under {\tt if}
conditionals.
If the program space is not closed under {\tt if} conditionals, or if 
we do not wish to use this form of unification for other reasons,
then the UNIF loop needs to be extended.
An example of a program space that is not closed under {\tt if}
conditionals,  
and that is an interesting synthesis target, are bit-vector
expressions. 

%
The STUN algorithm extends UNIF with backtracking (as explained in the
introduction, this
is needed since the unifcation can fail), and at each level,
a CEGIS loop can be used in {\tt syntFitSome}. 
The CEGIS and UNIF loops are thus combined, and the combination can be
fine-tuned for individual domains.

\vspace{-2ex}
\section{Synthesis through Unification algorithm}
\label{sec:algo}
\vspace{-1ex}

\comment{
\arsays{Need to talk about the following stuff:
  \begin{compactitem}
  \item Unification operator properties
  \item Specification form --- needs to be separable
  \end{compactitem}
}
}

\begin{algorithm}[t]
  \caption{The STUN (synthesis through unification) procedure}
  \label{algo:main}
  \renewcommand{\algorithmiccomment}[1]{\hfill //#1 }
  \renewcommand{\algorithmicrequire}{\textbf{Input:}}
  \renewcommand{\algorithmicensure}{\textbf{Output:}}
  \begin{algorithmic}[1]
    \REQUIRE Specification $\Spec$, Program space $\ProgSpace$, Input
    space $\InpSpace$, outer unification constraints (OUCs) $\UnifConstraints$
    \ENSURE $\Prog \in \ProgSpace$ s.t. $\forall \inp \in \InpSpace:
    \Prog[\inp] \models \Spec$ and $\Prog \models \UnifConstraints$,
    or $\None$
  \renewcommand{\algorithmicensure}{\textbf{Global variables:}}
  \ENSURE  learned unification constraints (LUCs) $\BackConstraints$,
  initialized to {\bf true}
  \renewcommand{\algorithmicensure}{\textbf{Output:}}
    \STATE $\constraints \gets {\tt true}$  \COMMENT { 
      CEGIS constraints } 
      \STATE \textbf{if}~{ $\InpSpace = \emptyset$ }
      \textbf{return}~$\top, \True$   \COMMENT { input space is empty,
        base case of recursion} \label{ln:basecase}
%
    \WHILE {$\True$} \label{ln:mainloop}
      \STATE $(\Prog,\TimeOut) \gets Generate(\ProgSpace,
      \Spec, \InpSpace, \constraints, \UnifConstraints,
      \BackConstraints)$ \COMMENT { 
                        generate next candidate} \label{ln:generate}
      \IF { $\Prog = \None$ } 
         \IF { $\neg \TimeOut$ }
           \STATE $\BackConstraints \gets \BackConstraints \wedge
              \mathit{LearnFrom}(Spec,\UnifConstraints,\BackConstraints)$
              \COMMENT {learn unification constraints} \label{ln:nosoln} 
         \ENDIF
         \STATE \textbf{return} $\None$ \COMMENT{no solution
           exists} 
      \ENDIF
      \STATE $\inp \gets \mathit{PickInput}(\InpSpace,\Prog)$ \COMMENT{take
        a positive- or a counter-example} \label{ln:pickinput}
      \IF { $\Prog[\inp] \not\models \Spec$ } \label{ln:ctrex}
        \STATE $\constraints \gets \constraints \wedge
           \mathit{project}(\Spec, \inp)$ \COMMENT {get a constraint
             from a counter-example} \label{ln:cegisconstr}
      \ELSE  \label{ln:else}
        \STATE $\InpSpace_G, \InpSpace_B \gets
                  \splitInpSpace(\Spec, \Prog, \inp)$  
                  \COMMENT{ $ \InpSpace_G \subseteq \{ \inp' \mid
                    \Prog[\inp'] \models \Spec \} $ } \label{ln:split} 
                  \COMMENT { and $\inp \in \InpSpace_G$, so
                    $\InpSpace_B \subsetneq 
                  \InpSpace$ and we can make a recursive call}
        \STATE $\Prog' \gets
                   \mathit{STUN}(\Spec, \UnifConstraints \wedge 
                   \mathit{UnifConstr}(\InpSpace_G, \Prog), \ProgSpace, \InpSpace_B)$ 
                   \COMMENT {recursive call } \label{ln:rec}
        \STATE  \textbf{if}~{ $\Prog' \neq \None$ }~\textbf{return}
        $\InpSpace_G \cdot \Prog \oplus \InpSpace_B \cdot
        \Prog'$ \label{ln:unif} 
                 \COMMENT {return the {\em unified} program} 
      \ENDIF
    \ENDWHILE
  \end{algorithmic}
\end{algorithm}

\noindent {\bf Overview.}
The STUN procedure is presented in Algorithm~\ref{algo:main}. The
input to the algorithm consists of a specification 
$\Spec$, a program space $\ProgSpace$, input space $\InpSpace$, and
{\em outer unification constraints (OUCs)} $\UnifConstraints$. OUCs are
constraints on the program space which are needed if the 
synthesized program will need to be unified with an already created
program. 
The algorithm is implemented as a recursive (backtracking) procedure
STUN. At each level, a decision is tried: a candidate program that
satisfies OUCs is generated, and passed to the recursive call. If the
recursive call is successful, the returned program is unified with the
current candidate. If the recursive call is unsuccessful, it records
{\em learned unification constraint (LUCs)} to the global variable
$\BackConstraints$, ensuring progress. 

\noindent {\bf Algorithm description.}
The algorithm first checks whether the input space is empty (this is
the base case of our recursion). If so, we return a program $\top$
(Line~\ref{ln:basecase}), a program which can be unified with any
other program.  

If the input space $\InpSpace$ is not empty, we start the main loop
(Line~\ref{ln:mainloop}). In the loop, we need
to generate a program $\Prog$ (Line~\ref{ln:generate}) that works for a
nonempty subset of $\InpSpace$. The generated program has to satisfy
``CEGIS'' constraints $\constraints$ (that ensure that the
program is correct on previously seen inputs at this level of
recursion), OUCs $\UnifConstraints$ that
ensure that the program  is unifiable with programs already created in
the upper levels of recursion, and LUCs $\BackConstraints$, which
collects constraints learned from the lower levels of recursion. 
If the call to {\tt Generate} fails (i.e., returns
$\None$), we exit this level of recursion, and learn constraints
unification constraints that can be inferred from the failed
exploration (Line~\ref{ln:nosoln}). The only exception is when {\tt
  Generate} fails due to a timeout, in which case we are not sure
whether the task was unrealizable, and so no constraints are learned.  
Learning the constraints (computed by the function
$\mathit{LearnFrom}$) is a form of conflict-driven learning. 

Once a program $\Prog$ is generated, we need to check whether it works for all
inputs in $\InpSpace$. 
If it does not, we need to decide
whether to improve $\Prog$ (in a CEGIS-like way), or generate a program
$\Prog'$ that works for inputs on which $\Prog$ does not work. 
The decision is made as follows. We pick an input
$\inp$ and check whether the program $\Prog$ is correct on $\inp$
(Line~\ref{ln:ctrex}). 
If $\Prog$ is not correct on $\inp$, then we have found a
counterexample, and we use it to strengthen our CEGIS constraints
(Line~\ref{ln:cegisconstr}). We refer to this branch as CEGIS-like
branch. 

If $\Prog$ is correct on $\inp$, then we know that $\Prog$ is
correct for at least one input, and we can make a recursive call to
generate a program that is correct for the inputs for which $\Prog$ is
not. We refer to this branch as the UNIF-like branch. The first step is
to split the input space $\InpSpace$ into the 
set $\InpSpace_G$ (an underapproximation of the set of inputs on which
$\Prog$ works containing at least $\inp$), and $\InpSpace_B$, the rest
of the inputs (Line~\ref{ln:split}).  
We can now make the recursive call on $\InpSpace_B$
(Line~\ref{ln:rec}). We pass the OUCs $\UnifConstraints$ to the
recursive call, in addition to the information that the returned
program will need to be unified with $\Prog$ (this is accomplished by
adding $\mathit{UnifConstr}(\InpSpace_G, \Prog)$). If the recursive
call does not 
find a program (i.e., returns $\Prog'=\None$), then the loop
continues, and another candidate is generated. If the recursive call
successfully returns a program $\Prog'$, this program is unified with 
with $\Prog$ (Line~\ref{ln:unif}). In more detail, we have a program
$\Prog$ that works on inputs in $\InpSpace_G$, and a program
$\Prog'$ that works on inputs in $\InpSpace_B$, and we unify them with
the unification operator $\oplus$ to produce $\InpSpace_G \cdot \Prog
\oplus \InpSpace_B \cdot \Prog'$. 
We know that the unification
operator will succeed, as the unification constraint
$\mathit{UnifConstr}(\InpSpace_G, \Prog)$ was passed to the recursive
call.

The input choice (line~\ref{ln:pickinput}), here 
nondeterministic, can be tuned for individual
domains to favor positive- or counter-examples, and
hence, CEGIS or UNIF. 

\noindent{\em Example 2.}
\addtocounter{example}{1}
Consider a specification that requires that the right-most bit 
set to $1$ in the input bit-vector is reset to $0$. This problem comes
from the Hacker's Delight collection~\cite{hacker}. 
A correct solution is, for instance, given by the expression
 $x \mathbin{\&} (x - 1)$.   
We illustrate the STUN procedure on this example.
The full STUN procedure for the bit-vector domain will be presented in
Section~\ref{sec:bv}.

\paragraph{Unification.} 
The unification operator $\InpSpace_G \cdot \Prog \oplus \InpSpace_B \cdot \Prog'$
works as follows. $\InpSpace_G \cdot \Prog$ and $\InpSpace_B \cdot \Prog'$ can be
unified if there exists a way to substitute the constants $c_i$ and
$c'_i$ occuring in $\Prog$ and $\Prog'$ with sub-expressions $\expr_i$
and $\expr'_i$ such that after the substitution, $\Prog$ and
$\Prog'$ are equal to the same program 
$\Prog^*$, and for all input in $\InpSpace_G$, $\expr_i[i] = c_i$ and
for all inputs in $\InpSpace_B$, $\expr_i'[i] = c_i'$. 
Note that this is a (very) simplified version of the unification
operator introduced in the next section. It is used here to illustrate
the algorithm. 

\paragraph{Unification gone wrong.}
Let us assume that the {\tt Generate} function at
Line~\ref{ln:generate} generates the program 
$x \mathbin{|} 0$ (this can happen if say the simpler programs already
failed). Note that $|$ is the bitwise or operator. 
Now let us assume that at Line~\ref{ln:pickinput}, we pick the input
$0$. The program matches $\Spec$ at this input. The set $\InpSpace_G$
is $\{0\}$, and we go to the recursive call at
Line~\ref{ln:rec} for the rest of the input space, with the constraint
that the returned program must be unifiable with
$x \mathbin{|} 0$. 
In the recursive call, {\tt Generate} is supposed to 
find a program that is unifiable with $x \mathbin{|} 0$, i.e., of the
form $x \mathbin{|} c$ for some constant $c$. 
Further, for the recursive call to finally succeed (i.e., take the else
branch at Line~\ref{ln:else}), we need this program to be correct on
some input other than $x = 0$. 
However, as it can be seen, there is no such program and input.
Hence, the procedure eventually backtracks while adding a constraint
that enforces that the program $x \mathbin{|} 0$ will no longer be attempted. 

\paragraph{Unification gone right.}
After the backtracking, with the additional constraint, the
program generation procedure is forbidden from generating the program 
$x \mathbin{|} 0$. The {\tt Generate} procedure instead generates say $x
\mathbin{\&} {-1}$. 
As before, for the recursive call to finally succeed, the program
generation procedure is asked to find a program unifiable with 
$x \mathbin{\&} {-1}$ 
(i.e., of the form $x \mathbin{\&} c$) that works for an input other than $0$.
Let us assume that generated program in the next level of recursion is
$x \mathbin{\&} 4$; one input for which this is correct is $x = 5$.
Attempting to unify these functions, the unification operator is asked
to find an expression $\expr$ such that $\expr[0/x] = {-1}$ and
$\expr[5/x] = 4$. 
One such candidate for $\expr$ is $x - 1$. 
This leads to a  valid solution $x \mathbin{\&} (x - 1)$ to the original synthesis
problem. 

\paragraph{Soundness.}
The procedure $\splitInpSpace(\Spec,\Prog,\inp)$ is sound if for every
invocation, it returns a pair $(\InpSpace_G, \InpSpace_B)$ such that 
$\{ \inp \} \subseteq \InpSpace_G \subseteq \{ \inp' \mid
\Prog[\inp'] \models \Spec \} \wedge \InpSpace_B = \InpSpace \setminus
\InpSpace_G$. 
The unification operator $\oplus$ is sound w.r.t. $\Spec$ and
$\ProgSpace$ if for programs $\Prog_1$ and $\Prog_2$ satisfying $\Spec$ on
inputs in $\InpSpace_1$ and $\InpSpace_2$, respectively, the program
$\InpSpace_1 \cdot \Prog_1 \oplus \InpSpace_2 \cdot \Prog_2$ is in
$\ProgSpace$ and that it satisfies $\Spec$ on $\InpSpace_1 \cup
\InpSpace_2$.
%
The procedure STUN is {\em sound} if for all inputs 
$\ProgSpace$, $\InpSpace$, $\Spec$, $\UnifConstraints$, it returns  a
program $\Prog$ such that $\Prog \in \ProgSpace$ and that $\forall
\inp \in \InpSpace: \Prog[\inp] \models \Spec$.

\vspace{-1ex}
\begin{theorem}\label{thm:ref}
Let us fix specification $\Spec$ and program space $\ProgSpace$. 
If $\splitInpSpace$ and 
the unification operator $\oplus$ are sound, then the STUN procedure is sound.
\end{theorem}
\vspace{-1ex}

\noindent{\bf Domains and Specifications.}
We instantiate STUN approach to three
domains: bit-vector expressions, separable specifications for conditional
linear-arithmetic 
expressions, and non-separable specifications for conditional linear
arithmetic expressions. 
Separable specifications are those where the specification relates an
input and its corresponding output, but does not constrain outputs that
correspond to different inputs.
Formally, we define separable specifications syntactically --- they
are of the form $f(x) = \OutputVariable \wedge \Phi(\OutputVariable,x)$,
where $x$ is the tuple of all input variables, $\OutputVariable$ is the
output variable, $f$ is the function being specified, and $\Phi$ is a
formula.
For example, the specification $\Spec \equiv f(x, y) \geq x \wedge f(x,
y) \geq y$ is separable as $\Spec = (f(x,y) = o) \wedge (o \geq x \wedge
o \geq y)$, and the specification $f(0) = 1 \vee f(1) = 1$ 
is a non-separable specification.

\noindent{\bf Notes about implementation.}
We have implemented the STUN procedure for each of the three domains
described above is a suite of tools.
In each case, we evaluate our tool on the benchmarks from the SyGuS
competition 2014~\cite{syguscomp}, and compare the performance of our
tool against the  
enumerative solver {\esolver}~\cite{sygus,AbhishelPLDI2013}.
The tool {\esolver} was the overall winner in the SyGuS competition
2014, and hence, is a good yardstick that represents the state of the
art. 

\vspace{-2ex}
\section{Domain: Bit-Vector Expressions}
\label{sec:bv}
\vspace{-1ex}

The first domain to which we apply the STUN approach is the domain of
bit-vector expressions specified by separable specifications.
Each bit-vector expression is either an input variable, a constant, or
a standard bit-vector operator applied to two sub-expressions.
This syntax does not have a top level {\tt if-then-else} operator that
allows unification of any two arbitrary programs.

Here, we instantiate the {\tt Generate} procedure and the unification
operator of Algorithm~\ref{algo:main} to obtain a nondeterministic
synthesis procedure (nondeterministic mainly in picking inputs that 
choose between the CEGIS-like and UNIF-like branches).
Later, we present a practical deterministic version of the algorithm.

\noindent{\bf Representing candidate programs.}
In the following discussion, we represent programs using an alternative
formalism that lets us lazily instantiate constants in the program.
This representation is for convenience only---the procedure
can be stated without using it.
Formally, a {\em candidate bit-vector program} $\Prog$ over inputs
$v_1,\ldots,v_n$ is a tuple $\langle \expr, \concreteConstraint \rangle$
where:
\begin{inparaenum}[(a)]
\item $\expr$ is a bit-vector expression over $\{ v_1, \ldots,
  v_n \}$ and auxiliary variables $\{ \SubProg_0, \ldots,
  \SubProg_m \}$ such that each $\SubProg_i$ occurs exactly once in
  $\expr$; and 
\item $\concreteConstraint$ is a satisfiable constraint over
  $\SubProg_i$'s.
\end{inparaenum}
Variables $\SubProg_i$ represent constants of $\expr$ whose exact values
are yet to be synthesized, and $\concreteConstraint$ is a constraint on
their values.
Intuitively, in the intermediate steps of the algorithm, instead of
generating programs with explicit constants, we generate programs with
symbolic constants along with constraints on them.
A concrete program can be obtained by replacing the symbolic constants
with values from some satisfying assignment of $\concreteConstraint$.

\noindent{\bf Unification.}
As mentioned briefly in Section~\ref{sec:algo}, two candidate programs
are unifiable if the constants occurring in the expressions can be
substituted with sub-expressions to obtain a common expression.
However, the presence of symbolic constants requires a more involved
definition of the unification operator. 
Further, note that the symbolic constants in the two programs do not
have to be the same.
Formally, programs $\Prog = \langle \expr, \comment{ (\SubProg_1, \ldots,
\SubProg_m),} \concreteConstraint \rangle$ and $\Prog' = \langle
\expr', \comment{(\SubProg_1', \ldots, \SubProg_m'),} \concreteConstraint'
\rangle$ over input spaces $\InpSpace$ and $\InpSpace'$ are unifiable
if:
\begin{compactitem}
\item There exists an expression $\expr^*$ that can be obtained from 
  $\expr$ by replacing each variable $\SubProg_i$ in $\expr$ by an
  expression $\expr_i$, over the formal inputs $\{ v_1, \ldots, v_n \}$
  and new auxiliary variables $\{ \SubProg_1^*, \ldots, \SubProg_k^* \}$.
  Further, the same expression $\expr^*$ should also be obtainable from
  $\expr'$ by replacing each of its sub-programs $\SubProg_i'$ by an
  expression $\expr_i'$.
\item Constraint $\concreteConstraint^* = \bigwedge_{\Valuation}
  \concreteConstraint[\forall i.  \expr_i[\Valuation] / \SubProg_i] \wedge
  \bigwedge_{\Valuation'} \concreteConstraint'[\forall i. 
  \expr_i'[\Valuation'] / \SubProg_i']$ is satisfiable.
  Here, $\Valuation$ and $\Valuation'$ range over inputs from
  $\InpSpace$ and $\InpSpace'$, respectively.
\end{compactitem}
If the above conditions hold, one possible unified program
$\InpSpace\cdot\Prog \oplus \InpSpace'\cdot \Prog'$ is $\Prog^*
= (\expr^*, \concreteConstraint^*)$.
Intuitively, in the unified program, each $\SubProg_i$ is replaced with
a sub-expression $\expr_i$, and further, $\concreteConstraint^*$ ensures
that the constraints from the individual programs on the value of these
sub-expressions are satisfied.

\begin{example}
  The programs $\Prog = (x~\&~\SubProg_0, \SubProg_0 =
  -1)$ and $\Prog' = (x~\&~\SubProg_0', \SubProg_0' = 4)$ over the
  input spaces $\InpSpace = (x = 0)$ and $\InpSpace' = (x = 5)$ can be
  unified into $(x~\&~(x - \SubProg_0^*), (0 - \SubProg_0^* = -1)
  \wedge (5 -\SubProg_0^* = 4))$.
  Here, both $\SubProg_0$ and $\SubProg_0'$ are replaced with $x -
  \SubProg_0^*$ and the constraints have been
  instantiated with inputs from corresponding input spaces.
\end{example}

\noindent{\bf Unification constraints.}
In this domain, an outer unification constraint $\UnifConstraints$ is
given by a candidate program $\Prog_T$.
Program $(\expr, \concreteConstraint) \models \UnifConstraints$ if
$\Prog_T = (\expr_T, \concreteConstraint_T)$ and $\expr$ can be
obtained from $\expr_T$ by replacing each $\SubProg_i^T$ with
appropriate sub-expressions.
A learned unification constraint $\BackConstraints$ is given by
$\bigwedge \Not(\Prog_F^i)$. 
Program $(\expr, \concreteConstraint) \models \BackConstraints$ if
for each $\Prog_F^i = (\expr_F, \comment {(\SubProg_1^F, \ldots),}
\concreteConstraint_F)$, there is no substitution of $\SubProg_i^F$'s
that transforms $\expr_F$ to $\expr$.
Intuitively, a $\Prog$ satisfies $\UnifConstraints = \Prog_T$ and
$\BackConstraints = \bigwedge \Not(\Prog_F^i)$ if $\Prog$ can be unified
with $\Prog_T$ and cannot be unified with any of $\Prog_F^i$.
Boolean combinations of unification constraints can be easily defined.
In Algorithm~\ref{algo:main}, we define $\mathit{UnifConstr}(\InpSpace, \Prog) =
\Prog$ and $\mathit{LearnFrom}(\Spec, \UnifConstraints, \BackConstraints) =
\Not(\UnifConstraints)$.
Note that using the alternate representation for programs having
symbolic constants lets us have a very simple $\mathit{LearnFrom}$ that
just negates $\UnifConstraints$ -- in general, a more complex
$\mathit{LearnFrom}$ might be needed.

\noindent{\bf Program generation.}
A simple $Generate$ procedure enumerates programs, ordered by
size, and checks if the expression satisfies all the constraints.

\begin{theorem}
  \label{thm:bv}
  Let $\mathbb{P}$ be Algorithm~\ref{algo:main} instantiated with
  the procedures detailed above.
  A procedure that executes the non-deterministic branches of
  $\mathbb{P}$ in a dove-tailed fashion 
  is a sound synthesis algorithm for bit-vector expressions specified by
  separable constraints.
  Further, if a solution exists, the procedure returns one.
\end{theorem}

\paragraph{A practical algorithm.}
We instantiate the non-deterministic choices in the procedure from
Theorem~\ref{thm:bv} to obtain a deterministic procedure.
Intuitively, this procedure maintains a set of candidate
programs and explores them in a fixed order based on size.
Further, we optimize the program generation procedure to only examine
programs that satisfy the unification constraints, instead of
following a generate-and-test procedure.
Additionally, we eliminate the recursive call in 
Algorithm~\ref{algo:main}, and instead store the variables
$\InpSpace_G$ locally with individual candidate programs.
Essentially, we pass additional information to convert the
recursive call into a tail call.
Formally, we replace $\concreteConstraint$ in the candidate programs
with $\{ (\Valuation_0, \concreteConstraint_0), \ldots, (\Valuation_k,
\concreteConstraint_k) \}$ where $\Valuation_i$'s are input valuations
that represent $\InpSpace_G$ from previous recursive calls.
Initially, the list of candidate programs contains the program $(
\SubProg_0, \emptyset)$.
In each step, we pick the first candidate (say $(\expr, \{
(\Valuation_0, \concreteConstraint_0), \ldots \})$) and
concretize $\expr$ to $\expr^*$ by substituting $\SubProg_i$'s
with values from a model of $\bigwedge_i \concreteConstraint_i$.
If $\expr^*$ satisfies $\Spec$, we return it.

\begin{algorithm}
  \caption{A deterministic STUN algorithm for bit-vector expressions\label{algo:bv_practical}}
  \begin{algorithmic}[1]
    \STATE $\mathit{Candidates} \gets \langle (\SubProg_0, \emptyset ) \rangle$
    \WHILE { $\True$ }
    \STATE 
    $(\expr, \{ (\Valuation_0, \concreteConstraint_0), \ldots \}) \gets \mathit{Candidates}[0]$
    \STATE $\expr^* \gets substitute(\expr, \mathit{getModel}(\bigwedge_i \concreteConstraint_i))$ \label{line:expr_star}
    \STATE \textbf{if} { $\not \exists \inp : \expr^*[\inp] \not\models \Spec$ }~\textbf{return}~$\expr^*$
    \STATE $\concreteConstraint_\inp \gets concretize(\expr, \Spec, \inp)$~where~$\expr^*[\inp] \not\models \Spec$
    \IF { $\neg \mathit{Satisfiable}(\concreteConstraint_\inp)$ }
    \STATE $\mathit{Candidates} \gets tail(\mathit{Candidates})$ //{Eliminate progs needing unif. with curr}\label{line:eliminate}
    \ELSE
    \STATE $\mathit{Candidates}[0] \gets (\expr,
    \{(\Valuation_0, \concreteConstraint_0) \ldots \} \cup \{
    (\Valuation_\inp, \concreteConstraint_\inp) \})$
    \IF { $\neg \mathit{Satisfiable}(\bigwedge \concreteConstraint_i
    \wedge \concreteConstraint_\inp)$ }
    \STATE $\mathit{Candidates} \gets tail(\mathit{Candidates})$
    \FORALL { $\SubProg_i \in \{ \SubProg_0 \ldots \}$, 
     $\expr' \gets LevelOneExpressions()$ }
    \STATE $\mathit{Candidates} \gets
    \mathit{append}(\mathit{Candidates}, \mathit{substitute}(\Prog,
    (\SubProg_i, \expr'))) $ \label{line:deepen}
    \ENDFOR
    \ENDIF
    \ENDIF
    \ENDWHILE
  \end{algorithmic}
\end{algorithm}

Otherwise, there exists an input $\inp$ on which $\expr^*$ is
  incorrect.
  We obtain a new constraint $\concreteConstraint_\inp$ on
  $\SubProg_i$'s by substituting the input and the expression $\expr^*$
  in the specification $\Spec$.
  If $\concreteConstraint_\inp$ is unsatisfiable, there are no expressions
  which can be substituted for $\SubProg_i$'s to make $\expr$ correct on
  $\inp$.
  Hence, the current candidate is eliminated--this is equivalent to a
  failing recursive call in the non-deterministic version.

Instead, if $\concreteConstraint_\inp$ is satisfiable,
  it is added to the candidate program.
  Now, if $\bigwedge \concreteConstraint_i \wedge
  \concreteConstraint_\inp$ is unsatisfiable, the
  symbolic constants $\SubProg_i$'s cannot be instantiated with explicit
  constants to make $\expr$ correct on all the seen inputs $\Valuation_i$.
  However, $\SubProg_i$'s can possibly be instantiated with other
  sub-expressions.
  Hence, we replace the current candidate with programs where each
  $\SubProg_i$ is replaced with a small expression of the form
  $operator(e_1, e_2)$ where $e_1$ and $e_2$ are either input variables
  or fresh $\SubProg_i$ variables.
  Note that while substituting these expression for $\SubProg_i$
  in $\concreteConstraint_j$, the input variables are replaced with the
  corresponding values from $\Valuation_j$.

Informally, each $(\expr, \concreteConstraint_i)$ is a candidate
program generated at one level of the recursion in the non-deterministic
algorithm and each valuation $\Valuation_i$ is the corresponding
input-space.
An iteration where $\concreteConstraint_\inp$ is unsatisfiable is a case
where there is no program that is correct on $\inp$ is unifiable with
the already generated program, and an iteration where $\bigwedge
\concreteConstraint_i \wedge \concreteConstraint_\inp$ is unsatisfiable
when the unification procedure cannot replace the symbolic
constants with explicit constants, but instead has to search through
more complex expressions for the substitution.

\noindent{\bf Experiments.}
We implemented Algorithm~\ref{algo:bv_practical} in a tool called
{\AUK} and evaluated it on benchmarks from the bit-vector track 
of SyGuS competition 2014~\cite{syguscomp}.
\iffull
As a representative subset of results, we present the running times on
the $59$ hacker's delight benchmarks in the appendix
(Table~\ref{table:bv_results}).
\fi
For easy benchmarks (where both tools take $< 1$ second),
{\esolver} is faster than {\AUK}.
However, on larger benchmarks, the performance of {\AUK} is better.
We believe that these results are due to {\esolver} being able to
enumerate small solutions extremely fast, while {\AUK} starts on the
expensive theory reasoning.
On larger benchmarks, {\AUK} is able to eliminate larger sets of
candidates due to the unification constraints while {\esolver} is
slowed down by the sheer number of candidate programs.

\vspace{-2ex}
\section{Domain: CLEs with Separable Specifications}
\label{sec:lia}
\vspace{-1ex}

We now apply the STUN approach to the domain of conditional
linear arithmetic expressions (CLEs).
A program $\Prog$ in this domain is either a linear expression over the
input variables or is ${\tt if(cond)~\Prog~else~\Prog'}$, where {\tt cond}
is a boolean combination of linear inequalities.
This is an ideal domain for the UNIF loop 
due to the natural unification operator that uses the
\textsf{if-then-else} construct.
Here, we present our algorithm for the case where the variables range
over rationals.
Later, we discuss briefly how to extend the technique to integer
variables.

\noindent{\bf Unification.}
Given two CLEs $\Prog$ and $\Prog'$,
and input spaces $\InpSpace$ and $\InpSpace'$, we define $\InpSpace
\cdot \Prog \oplus \InpSpace' \cdot \Prog'$ to be the program
$\mathsf{if}~(\InpSpace)~\Prog~\mathsf{else~if}~(\InpSpace')~\Prog'~\mathsf{else}~\bot$.
Note that we assume that $\InpSpace$ and $\InpSpace'$ are expressed as
linear constraints.
Here, since any two programs can be unified,
unification constraints are not used.  

\noindent{\bf Program Generation.}
Algorithm~\ref{algo:lra_generate} is the program generation procedure
$\mathit{Generate}$ for CLEs for rational
arithmetic specifications.
Given a specification $\Spec$ and input space $\InpSpace$, it first
generates a concrete input-output example such that the input is in
$\InpSpace$ and example satisfies $\Spec$.
Then, it generalizes the input-output pair to a program as follows. 
From each clause of the specification $\Spec$, we pick one disjunct that
evaluates to true for the current input-output pair.
Each disjunct that constrains the output can be expressed as
$\OutputVariable~\mathsf{op}~\phi$ where $\mathsf{op} \in \{ {\leq},
{\geq}, {<}, {>} \}$ and $\phi$ is a linear expression over the input
variables.
Recall from the definition of separable specifications that
$\OutputVariable$ is the output variable that represents the output of
the function to be synthesized.
Each such inequality gives us a bound on the output variable.
The algorithm then returns an expression $\Prog$ that respects the
strictest (in the input-output example) bounds among these.
Further, we define the $\mathit{SplitInpSpace}$ procedure from
Algorithm~\ref{algo:main} as follows: the input space $\InpSpace_G$ is
obtained by substituting the program $\Prog$ into the disjuncts.
The space $\InpSpace_B$ is defined as $\InpSpace \wedge \neg
\InpSpace_G$.

\begin{algorithm}
  \caption{Procedure $\mathit{Generate}$\label{algo:lra_generate}}
  \begin{algorithmic}[1]
    \REQUIRE Specification $\Spec$ in CNF,  Input space $\InpSpace$
    \ENSURE Candidate program $\Prog$
    \STATE \textbf{if}~{$\InpSpace = \emptyset$}~\textbf{return}~$\top$
    \STATE $\mathit{pex} \gets \mathit{getModel}(\InpSpace \wedge \Spec)$
    \STATE $\mathit{LB} \gets -\infty$, $\mathit{UB} \gets
    \infty$
    \FORALL {$\mathit{Clause}$ of $\Spec$}
    \STATE {Pick $\mathit{Disjunct}$ in $\mathit{Clause}$ such that
    $\mathit{Disjunct}[\mathit{pex}]$ holds}
    \IF {$\OutputVariable$ occurs in $\mathit{Disjunct}$ and $
    \mathit{Disjunct}\equiv(\OutputVariable~\mathsf{op}~\phi)$}
    \STATE $\quad\mathbf{case}~\mathsf{op} \in \{ \leq, < \} \wedge
    \mathit{UB}[\mathit{pex}] > \phi[\mathit{pex}]\quad:\quad~\mathit{UB} \gets \phi$
    \STATE $\quad\mathbf{case}~\mathsf{op} \in \{ \geq, > \} \wedge
    \mathit{LB}[\mathit{pex}] < \phi[\mathit{pex}]\quad:\quad~\mathit{LB} \gets \phi$
    \ENDIF
    \ENDFOR
    \RETURN $(\mathit{LB} + \mathit{UB}) / 2$
  \end{algorithmic}
\end{algorithm}
\vspace{-2ex}
\begin{theorem}
  Algorithm~\ref{algo:main} instantiated with the generation and
  unification procedures detailed above is a sound and complete
  synthesis procedure for conditional linear rational arithmetic
  expressions specified using separable specifications.
\vspace{-1ex}
\end{theorem}

\noindent{\bf Extension to integers.}
The above procedure cannot be directly applied when variables range over
integers instead of rationals.
Here, each disjunct can be put into the form $c \cdot
\OutputVariable~\mathsf{op}~\phi$ where $c$ is a positive integer
and $\phi$ is a linear integer expression over inputs.
For rationals, this constraint can be normalized to obtain
$\OutputVariable~\mathsf{op}~\frac{1}{c} \phi$.
In the domain of integers, $\frac{1}{c} \phi$ is not necessarily an
integer.

There are two possible ways to solve this problem. 
A simple solution is to modify the syntax of the programs to
allow floor $\lfloor \cdot \rfloor$ and ceiling $\lceil \cdot \rceil$
functions.
Then, $c \cdot \OutputVariable \leq \phi$ and $c \cdot \OutputVariable
\geq \phi$ can be normalized as $\OutputVariable \leq \lfloor \phi / c
\rfloor$ and $\OutputVariable \geq \lceil \phi/ c \rceil$.
The generation procedure can then proceed using these normalized
expressions.
The alternative approach is to use a full-fledged decision procedure for
solving the constraints of the form
$\OutputVariable~\mathsf{op}~\frac{1}{c} \phi$.
However, this introduces divisibility constraints into the generated
program.
For a detailed explanation on this approach and techniques for
eliminating the divisibility constraints, see~\cite{Comfusy}.

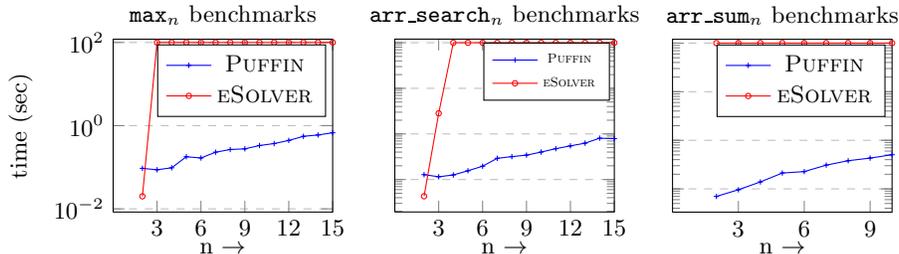
\begin{figure}
    \pgfplotsset{width=4.5cm} 
    \begin{tikzpicture}
      \begin{semilogyaxis}[
          title={$\mathtt{max}_n$ benchmarks},
          xlabel={n $\to$},
          ylabel={time (sec)},
          xmin=0, xmax=15,
          ymin=0, ymax=120,
          xtick={3,6,9,12,15},
          legend pos=north east,
          ymajorgrids=true,
          grid style=dashed,
          title style={yshift=-1.0ex,},
          x label style={at={(axis description cs:0.5,+0.1)}},
        ]

        \addplot[
          color=blue,
          mark=+,
          mark size=1pt
        ]
        coordinates {
          (2, 0.094)
          (3, 0.087)
          (4, 0.097)
          (5, 0.179)
          (6, 0.167)
          (7, 0.230)
          (8, 0.267)
          (9, 0.277)
          (10, 0.333)
          (11, 0.371)
          (12, 0.441)
          (13, 0.554)
          (14, 0.597)
          (15, 0.675)
        };
        \addplot[
          color=red,
          mark=o,
          mark size=1pt
        ]
        coordinates {
          (2, 0.020)
          (3, 100)
          (4, 100)
          (5, 100)
          (6, 100)
          (7, 100)
          (8, 100)
          (9, 100)
          (10, 100)
          (11, 100)
          (12, 100)
          (13, 100)
          (14, 100)
          (15, 100)
        };
        \legend{{\PUFFIN}, {\esolver}}

      \end{semilogyaxis}
    \end{tikzpicture}
    \begin{tikzpicture}
      \begin{semilogyaxis}[
          title={$\mathtt{arr\_search}_n$ benchmarks},
          xlabel={n $\to$},
          xmin=0, xmax=15,
          ymin=0, ymax=120,
          xtick={3,6,9,12,15},
          yticklabels={,,}
          legend pos=north east,
          ymajorgrids=true,
          grid style=dashed,
          title style={yshift=-1.0ex,},
          x label style={at={(axis description cs:0.5,+0.1)}},
        ]

        \addplot[
          color=blue,
          mark=+,
          mark size=1pt
        ]
        coordinates {
          (2, 0.128)
          (3, 0.115)
          (4, 0.126)
          (5, 0.156)
          (6, 0.196)
          (7, 0.292)
          (8, 0.318)
          (9, 0.343)
          (10, 0.400)
          (11, 0.475)
          (12, 0.547)
          (13, 0.627)
          (14, 0.807)
          (15, 0.789)
        };
        \addplot[
          color=red,
          mark=o,
          mark size=1pt
        ]
        coordinates {
          (2, 0.043)
          (3, 2.834)
          (4, 100)
          (5, 100)
          (6,100)
          (7,100)
          (8,100)
          (9,100)
          (10,100)
          (11,100)
          (12,100)
          (13,100)
          (14,100)
          (15,100)
        };
        \legend{\tiny{\PUFFIN}, \tiny{\esolver}}

      \end{semilogyaxis}
    \end{tikzpicture}
    \begin{tikzpicture}
      \begin{semilogyaxis}[
          title={$\mathtt{arr\_sum}_n$ benchmarks},
          xlabel={n $\to$},
          xmin=0, xmax=10,
          ymin=0, ymax=120,
          xtick={3,6,9,12,15},
          yticklabels={,,},
          legend pos=north east,
          ymajorgrids=true,
          grid style=dashed,
          title style={yshift=-1.0ex,},
          x label style={at={(axis description cs:0.5,+0.1)}},
        ]

        \addplot[
          color=blue,
          mark=+,
          mark size=1pt
        ]
        coordinates {
          (2, 0.070)
          (3, 0.096)
          (4, 0.139)
          (5, 0.212)
          (6, 0.227)
          (7, 0.308)
          (8, 0.380)
          (9, 0.430)
          (10, 0.505)
        };
        \addplot[
          color=red,
          mark=o,
          mark size=1pt
        ]
        coordinates {
          (2,100)
          (3,100)
          (4,100)
          (5,100)
          (6,100)
          (7,100)
          (8,100)
          (9,100)
          (10,100)
          (11,100)
          (12,100)
          (13,100)
          (14,100)
          (15,100)
        };
        \legend{{\PUFFIN}, {\esolver}}
      \end{semilogyaxis}
    \end{tikzpicture}
    \caption{Results on separable linear integer benchmarks \label{fig:lra_results} }
\end{figure}
\noindent{\bf Experiments.}
We implemented the above procedure in a tool called {\PUFFIN} 
and evaluated it on benchmarks from the linear
integer arithmetic track with separable specifications from the SyGuS
competition 2014.
The results on three classes of benchmarks ($\mathtt{max}_n$,
$\mathtt{array\_search}_n$, and $\mathtt{array\_sum}_n$) have been
summarized in Figure~\ref{fig:lra_results}.
The $\mathtt{max}_n$ benchmarks specify a function that outputs the
maximum of $n$ input variables (the illustrative example from
Section~\ref{sec:illustrative_example} is $\max_2$)\footnote{Note that
the SyGuS competition benchmarks only go up to $\max_5$}.
The $\mathtt{array\_search}_n$ and $\mathtt{array\_sum}_n$ benchmarks
respectively specify functions that search for a given input in a sorted
array, and check if the sum of two consecutive elements in an array 
is equal to a given value.
In all these benchmarks, our tool significantly outperforms {\esolver}
and other existing solvers based on a pure CEGIS approach.
The reason for this is as follows: the CEGIS solvers try to generate
the whole program at once, which is a complex expression. 
On the other hand, our solver combines simple expressions generated for
parts of the input spaces where the output expression is simple.

\vspace{-2ex}
\section{Domain: Non-Separable Specifications for CLEs}
\label{sec:lia_nonfunc}
\vspace{-1ex}

Here, we consider CLEs specified by non-separable specifications.
While this domain allows for simple unification, non-separable
specifications introduce complications.
Further, unlike the previous domains, the problem itself is
undecidable. 
%

First, we define what it means for a program $\Prog$ to satisfy a
non-separable specification on an input space $\InpSpace$.
We say that $\Prog$ satisfies $\Spec$ on $\InpSpace$ if $\Spec$ holds
whenever the inputs to the function in each invocation
in $\Spec$ belong to $\InpSpace$.
For example, program $\Prog(i)$ satisfies $\Spec \equiv f(x) = 1 \wedge
x' = x + 1 \implies f(x') = 1$ on the input space $0 \leq i \leq 2$ if
$(0 \leq x \leq 2 \wedge 0 \leq x' \leq 2) \implies \Spec[f \gets
\Prog]$ holds, i.e., we $\Spec$ to hold when both $x$ and $x'$  belong
to the input space.
In further discussion, we assume that the program to be synthesized is
represented by the function $f$ in all specifications and formulae.


\noindent{\bf Unification and Unification constraints.}
The unification operator we use is the same as in the previous
section.
%
However, for non-separable specifications, the outputs produced by
$\Prog$ on $\InpSpace$ may constrain the outputs of $\Prog'$ on
$\InpSpace'$, and hence, we need non-trivial unification constraints.
An outer unification constraint $\UnifConstraints$ is a sequence
$\langle (\InpSpace_0, \Prog_0), (\InpSpace_1, \Prog_1), \ldots \rangle$
where $\InpSpace_i$'s and $\Prog_i$'s are input spaces and programs,
respectively. A learned unification constraint
$\BackConstraints$ is given by $\bigwedge \concreteConstraint_i$ where
each $\concreteConstraint_i$ is a formula over $f$.
Intuitively, $\InpSpace_i$ and $\Prog_i$ fix parts of the synthesized
function, and the constraints $\concreteConstraint_i$ 
enforce the required relationships between the outputs produced by
different $\Prog_i$'s.
Formally, $\Prog \models \UnifConstraints$ if 
its outputs agree with each $\Prog_i$ on $\InpSpace_i$ and 
$\Prog \models \BackConstraints$ if $\wedge
\concreteConstraint_i[\Prog / f]$ holds.

\noindent{\bf Program Generation.}
The $\mathit{Generate}$ procedure works using input-output examples as
in the previous section.
However, it is significantly more complex due to the presence of
multiple function invocations in $\Spec$.
Intuitively, we replace all function invocations except one with the
partial programs from the unification constraints and then solve the
arising separable specification using techniques from the previous section.
We explain in detail using an example.
\begin{example}
  Consider the specification $\Spec$ given by $x \neq y \implies f(x) +
  f(y) = 10$.
  Here, the only solution is the constant function $5$.
  Now, assume that the synthesis procedure has guessed that $\Prog_0$
  given by $\Prog_0(i) = 0$ is a program that satisfies $\Spec$ for the
  input space $\InpSpace_0 \equiv i = 0$.

  The unification constraint $\UnifConstraints_0 = \langle (\Prog_0,
  \InpSpace_0) \rangle$ is passed to the recursive call to ensure
  that the synthesized function satisfies $f(0) = 0$.
  The program generation function in the recursive call works as
  follows: it replaces the invocation $f(x)$ in $\Spec$ with the partial
  function from $\UnifConstraints$ to obtain
  the constraint $(x = 0 \wedge x \neq y \implies
  \Prog_0(0) + f(y) = 10)$.
  Solving to obtain the next program and input space,
  we get $\Prog_1(i) = 10$ for the input space $\InpSpace_1 \equiv i = 1$.
  Now, the unification constraint passed to the next recursive call is 
  $\UnifConstraints = \langle (\Prog_0, \InpSpace_0), (\Prog_1,
  \InpSpace_1) \rangle$.

  Again, instantiating $f(x)$ with $\Prog_0$ and $\Prog_1$ in
  the respective input spaces, we obtain the constraint 
  $(x = 0 \wedge x \neq y \implies \Prog_0(x) + f(y) = 10) \wedge (x = 1
  \wedge x \neq y \implies \Prog_1(x) + f(y) = 10)$.
  Now, this constraint does not have a solution---for $y = 2$, there is
  no possible value for $f(y)$.
  Here, a reason $\BackConstraints = \concreteConstraint_0$ (say
  $\concreteConstraint \equiv f(1) = f(0)$) is learnt for the
  unsatisfiability and added to the learned constraint.
  Note that this conflict-driven learning is captured in the function
  $\mathit{LearnFrom}$ in Algorithm~\ref{algo:main}.
  Now, in the parent call, no program satisfies
  $\BackConstraints$ as well as $\UnifConstraints = \langle (\Prog_0,
  \InpSpace_0), (\Prog_1, \InpSpace_1) \rangle$.
  By a similar unsatisfiability analysis, we get $\concreteConstraint_1
  \equiv f(0) = 5$ as the additional learned constraint.
  Finally, at the top level, with $\BackConstraints \equiv f(0) = f(1)
  \wedge f(0) = 5$, we synthesize the right value for $f(0)$.
\end{example}

\begin{example}[Acceleration]
  Let $\Spec \equiv \left( 0 \leq x,y \leq 2 \implies  f(x, y) = 1 \right)
  \wedge (x = 4 \wedge y = 0  \implies  f(x, y) = 0 )
  \wedge (f(x, y) = 1 \wedge (x', y') = (x + 2, y + 2)
  \implies  f(x', y') = 1 )$.

  The synthesis procedure first obtains the candidate program
  $\Prog_0(i, j) = 1$ on the input space $\InpSpace_0 \equiv 0 \leq
  i \leq 1 \wedge 0 \leq j \leq 1$.
  The recursive call is passed $(\Prog_0, \InpSpace_0)$ as the
  unification constraint and generates the next program fragment $\Prog_1(i,
  j) = 1$ on the input space $\InpSpace_1 \equiv 0 \leq i - 2 \leq 2
  \wedge 0 \leq j - 2 \leq 2$.
  Similarly, each further recursive call generates 
  $\Prog_n(i, j) = 1$ on the input space $\InpSpace_n$ given by $0 \leq
  i - 2*n \leq 2$.
  The sequence of recursive calls do not terminate.

  To overcome this problem, we use an accelerating widening operator.
  Intuitively, it generalizes the programs and input spaces in
  the unification constraints to cover more inputs.
  In this case, the acceleration operator we define below produces the
  input space $\InpSpace^* \equiv 0 \leq i \wedge 0 \leq j \wedge -2
  \leq i - j \leq 2$.
  Proceeding with this widened constraint lets us terminate with
  the solution program.
\end{example}
\noindent{\bf Acceleration.}
The {\em accelerating widening operator} $\nabla$ operates on
unification constraints.
In Algorithm~\ref{algo:main}, we apply $\nabla$ to the unification
constraints being passed to the
recursive call on line~\ref{ln:rec}, i.e., we replace the expression
$\UnifConstraints \wedge \mathit{UnifConstr}(\InpSpace_G, \Prog)$ with 
$\nabla(\UnifConstraints \wedge \mathit{UnifConstr}(\InpSpace_G, \Prog),
\BackConstraints)$.

While sophisticated accelerating widening operators are available for
partial functions (see, for example,~\cite{CC12,CC91}), in our
implementation, we use a simple one.
Given an input unification constraint $\langle
(\InpSpace_0, \Prog_0), \ldots, (\InpSpace_n, \Prog_n)\rangle$, the
accelerating widening operator works as follows:
\begin{inparaenum}[(a)]
\item If $\Prog_n \neq \Prog_j$ for all $j < n$, it returns the input.
\item Otherwise, $\Prog_n = \Prog_j$ for some $j < n$ and we widen the
  domain where $\Prog_n$ is applicable to $\InpSpace^*$ where 
  $\InpSpace_j \cup \InpSpace_n \subseteq \InpSpace^*$.
  Intuitively, we do this by letting $\InpSpace^* = 
  \nabla(\InpSpace_i, \InpSpace_j)$  where $\nabla$ is the widening join
  operation for convex polyhedra abstract domain~\cite{CH78}.
  However, we additionally want $\Prog_n$ on $\InpSpace^*$ to not cause 
  any violation of the learned constraints $\BackConstraints = \bigwedge
  \concreteConstraint_i$.
  Therefore, we use a widening operator with bounds on the convex
  polyhedral abstract domain instead of the generic widening operator.
  The bounds are obtained from the concrete constraints.
  We do not describe this procedure explicitly, but present an example
  below.
  The final output returned is $\langle (\InpSpace_0, \Prog_0),
  \ldots, (\InpSpace^*, \Prog_n) \rangle$.
\end{inparaenum}

\begin{example}
  Consider the specification $\Spec = f(0) = 1 \wedge (f(x) = 1 \wedge
  0 \leq x \leq 10 \implies f(x+1) = 1) \wedge (f(12) = 0)$.
  After two recursive calls, we get
  the unification constraint $\UnifConstraints = \langle (i = 0,
  \Prog_0(i) = 1), (i = 1, \Prog_1(i) = 1) \rangle$.
  Widening, we generalize the input spaces $i =
  0$ and $i = 1$ to $\InpSpace^* = (i \geq 0)$.
  However, further synthesis fails due to the clause
  $f(12) = 0$ from $\Spec$, and we obtain a learned unification
  constraint $\BackConstraints \equiv f(12) = 0$ at the parent call.

  We then obtain an additional bound for the unification as replacing
  $f$ by $\Prog_1$ violates $f(12) = 0$.
  With this new bound, the widening operator returns the input space
  $\InpSpace^* = (12 > i \geq 0)$, which allows us to complete the
  synthesis.
\end{example}

\vspace{-2ex}
\begin{theorem}
  Algorithm~\ref{algo:main} instantiated with the procedures described
  above is a sound synthesis procedure for conditional linear
  expressions given by non-separable specifications. 
\end{theorem}
\vspace{-2ex}

\comment{
\begin{wrapfigure}{l}{0.38\textwidth}
  \vspace{-1ex}
    \pgfplotsset{width=4.5cm} 
    \begin{tikzpicture}
      \begin{loglogaxis}[
          xlabel={{\esolver} (sec)},
          ylabel={{\RAZORBILL} (sec)},
          xmin=0, xmax=120,
          ymin=0, ymax=120,
          ymajorgrids=true,
          grid style=dashed,
        ]

        \addplot[
          color=blue,
          mark=+,
          mark size=1pt
        ]
        coordinates {
          (2, 0.094)
          (3, 0.087)
          (4, 0.097)
          (5, 0.179)
          (6, 0.167)
          (7, 0.230)
          (8, 0.267)
          (9, 0.277)
          (10, 0.333)
          (11, 0.371)
          (12, 0.441)
          (13, 0.554)
          (14, 0.597)
          (15, 0.675)
        };
        \addplot[
          color=red,
          mark=o,
          mark size=1pt
        ]
        coordinates {
          (1,0.1)
        };
      \end{loglogaxis}
    \vspace{-2ex}
    \end{tikzpicture}
    \caption{Invgen benchmarks \label{fig:la_nr_results}}
    \vspace{-4ex}
\end{wrapfigure}
}
\noindent{\bf Experiments.}
We implemented the above procedure in a tool called {\RAZORBILL}
and evaluated it linear integer benchmarks with non-separable
specifications from SyGuS competition 2014.
\iffull
We compare the performance of our tool and {\esolver} on the $29$ invgen
set of benchmarks (results in Table~\ref{table:lia_nr_results} in the
appendix).
\fi
As for the bit-vector benchmarks, on small benchmarks
(where both tools finish in less than $1$ second),
{\esolver} is faster.
However, on larger benchmarks, {\RAZORBILL} can be much faster.
As before, we hypothesize that this is due to {\esolver}
quickly enumerating small solutions before the STUN based solver can
perform any complex theory reasoning.

\vspace{-2ex}
\section{Concluding Remarks}
\vspace{-1ex}

\paragraph{Related work.} 
Algorithmic program synthesis became popular a decade ago with the
introduction of CEGIS~\cite{asplos06}. Much more recently, syntax-guided
synthesis~\cite{sygus} framework, where the input to
synthesis is a program space and a specification, was introduced,
along with several types of solvers. Our 
synthesis problem falls into this framework, and our solvers
solve SyGuS problem instances.  
Kuncak et al.~\cite{Comfusy} present another alternative (non-CEGIS)
solver for linear arithmetic constraints. 

STUN is a general approach to synthesis. For
instance, in the domain of synthesis of
synchronization~\cite{bloem14,vechev10,CAV13,CAV14,POPL15}, the
algorithm used can be presented as an instantiation of STUN. The approach
is based  on an analysis of a counterexample trace that infers a fix
in the form of  
additional synchronization. The bug fix works for the counterexample and
possibly for some related traces. Such bug fixes are then unified
similarly as in the STUN approach.  

A synthesis technique related to STUN is based on version-space
algebras~\cite{Gulwani11,LauDW00}. There, the goal is to
compose programs that works on a part of a single input (say a string)
to a 
transformation that would work for the complete single input. 
In contrast, STUN unifies programs that work for different
parts of the input space. 
The combination of the two approaches could thus be fruitful. 

The widening operator has been introduced
in~\cite{CousotCousot77}, and has   
been widely used in program analysis, but not in synthesis. We
proposed to use it 
to accelerate the process in which 
STUN finds solutions that cover parts of the input space. Use of
other operators such as narrowing is worth investigating. 

\paragraph{Limitations.}
We mentioned that the simple unification operator based on if
statements might lead to inefficient code. In particular, if the
specification is given only by input-output examples, the resulting
program might be a sequence of conditionals with conditions
corresponding to each example. 
That is why we proposed a different unification operator for the
bit-vector domain, and we plan to investigate unification further. 
Furthermore, a limitation of STUN when compared to CEGIS is that
designing unification operators requires domain knowledge (knowledge
of the given program space). 


\paragraph{Future work.} 
We believe STUN opens several new directions for future research. 
First, we plan to investigate unification operators for domains where
the programs have loops or recursion. This seems a natural
fit for STUN, because if for several different input we find that the
length of the synthesized sequence of instructions in the solution
depends on the 
size of the input, then the unification operator might propose a loop
in the unified solution.  
Second,  
systems that at runtime prevent deadlocks or other problems can be
thought of as finding solutions for parts of the input space. A
number of such fixes could then be unified into a more general
solution. 
Last, we plan to optimize the prototype solvers we presented. This is
a promising direction, as even our current prototypes have comparable or
significantly better performance than the existing solvers.

\newpage
\bibliographystyle{splncs03}
\bibliography{references}

\iffull

\newpage

\begin{appendix}
  \section{Appendix to Section~\ref{sec:lia}}

  \begin{table}
    \begin{tabular}{|c|c|r|r|   |c|c|r|r|}
      \hline
      Sl. no & Benchmark        & {\PUFFIN}     & {\esolver}       &
      Sl. no & Benchmark        & {\PUFFIN}     & {\esolver}        \\
             &                  & {Time (sec)}  &  Time (sec)  &
      &                  & {Time (sec)}  &  Time (sec)    \\
      \hline
      \hline
      1   & $\mathtt{max}_{2 }$&   0.094       &  0.020        &         24   & $\mathtt{array\_search}_{11}$ & 0.475 & TO \\
      2   & $\mathtt{max}_{3 }$&   0.087       & TO   &                  25   & $\mathtt{array\_search}_{12}$ & 0.547 & TO \\
      3   & $\mathtt{max}_{4 }$&   0.097       & TO  &                   26   & $\mathtt{array\_search}_{13}$ & 0.627 & TO \\
      4   & $\mathtt{max}_{5 }$&   0.179       & TO  &                   27   & $\mathtt{array\_search}_{14}$ & 0.807 & TO \\
      5   & $\mathtt{max}_{6 }$&   0.167       & TO  &                   28   & $\mathtt{array\_search}_{15}$ & 0.789 & TO \\
      6   & $\mathtt{max}_{7 }$&   0.230       & TO  &                   29   & $\mathtt{array\_sum5}\_2 $    & 0.070 & TO \\
      7   & $\mathtt{max}_{8 }$&   0.267       & TO  &                   30   & $\mathtt{array\_sum5}\_3 $    & 0.096 & TO \\
      8   & $\mathtt{max}_{9 }$&   0.277       & TO  &                   31   & $\mathtt{array\_sum5}\_4 $    & 0.139 & TO \\
      9   & $\mathtt{max}_{10}$&   0.333       & TO  &                   32   & $\mathtt{array\_sum5}\_5 $    & 0.212 & TO \\
      10   & $\mathtt{max}_{11}$&   0.371       & TO  &                  33   & $\mathtt{array\_sum5}\_6 $    & 0.227 & TO \\
      11   & $\mathtt{max}_{12}$&   0.441       & TO  &                  34   & $\mathtt{array\_sum5}\_7 $    & 0.308 & TO \\
      12   & $\mathtt{max}_{13}$&   0.554       & TO  &                  35   & $\mathtt{array\_sum5}\_8 $    & 0.380 & TO \\
      13   & $\mathtt{max}_{14}$&   0.597       & TO  &                  36   & $\mathtt{array\_sum5}\_9 $    & 0.430 & TO \\
      14   & $\mathtt{max}_{15}$&   0.675       & TO  &                  37   & $\mathtt{array\_sum5}\_10$    & 0.505 & TO \\
      15   & $\mathtt{array\_search}_{2 }$ & 0.128 & 0.043  &            38   & $\mathtt{array\_sum15}\_2 $   & 0.094 & TO \\
      16   & $\mathtt{array\_search}_{3 }$ & 0.115 & 2.834  &            39   & $\mathtt{array\_sum15}\_3 $   & 0.103 & TO \\
      17   & $\mathtt{array\_search}_{4 }$ & 0.126 & TO  &               40   & $\mathtt{array\_sum15}\_4 $   & 0.164 & TO \\
      18   & $\mathtt{array\_search}_{5 }$ & 0.156 & TO  &               41   & $\mathtt{array\_sum15}\_5 $   & 0.207 & TO \\
      19   & $\mathtt{array\_search}_{6 }$ & 0.196 & TO  &               42   & $\mathtt{array\_sum15}\_6 $   & 0.228 & TO \\
      20   & $\mathtt{array\_search}_{7 }$ & 0.292 & TO  &               43   & $\mathtt{array\_sum15}\_7 $   & 0.284 & TO \\
      21   & $\mathtt{array\_search}_{8 }$ & 0.318 & TO  &               44   & $\mathtt{array\_sum15}\_8 $   & 0.363 & TO \\
      22   & $\mathtt{array\_search}_{9 }$ & 0.343 & TO  &               45   & $\mathtt{array\_sum15}\_9 $   & 0.445 & TO \\
      23   & $\mathtt{array\_search}_{10}$ & 0.400 & TO  &               46   & $\mathtt{array\_sum15}\_10$   & 0.546 & TO \\
      \hline
    \end{tabular}
    \caption{Results for Section~\ref{sec:lia}\label{table:lra_results}}
  \end{table}

We implemented Algorithm~\ref{algo:bv_practical} in a tool called
{\AUK} and evaluated it on benchmarks from the bit-vector track 
of SyGuS competition 2014~\cite{syguscomp}.
As a representative subset of results, we present the running times on
the $59$ hacker's delight benchmarks in the appendix
(Table~\ref{table:bv_results}).
For easy benchmarks (where both tools take $< 1$ second),
{\esolver} is faster than {\AUK}.
However, on larger benchmarks, the performance of {\AUK} is better.
We believe that these results are due to {\esolver} being able to
enumerate small solutions extremely fast, while {\AUK} starts on the
expensive theory reasoning.
On larger benchmarks, {\AUK} is able to eliminate larger sets of
candidates due to the unification constraints while {\esolver} is
slowed down by the sheer number of candidate programs.

  \newpage
  \section{Appendix to Section~\ref{sec:bv}}
  \begin{table}
    \begin{tabular}{|c|c|r|r|   |c|c|r|r|}
      \hline
      Sl. no & Benchmark        & {\AUK}     & {\esolver}       &
      Sl. no & Benchmark        & {\AUK}     & {\esolver}        \\
             &                  & {Time (sec)}  &  Time (sec)  &
      &                  & {Time (sec)}  &  Time (sec)    \\
      \hline
      \hline
      01 &  $\mathtt{hd{\mbox{-}}01{\mbox{-}}d1{\mbox{-}}prog}$  &  0.074   & 0.030 & 29 &  $\mathtt{hd{\mbox{-}}10{\mbox{-}}d5{\mbox{-}}prog}$  &  0.888   & 0.406 \\
      02 &  $\mathtt{hd{\mbox{-}}01{\mbox{-}}d5{\mbox{-}}prog}$  &  0.106   & 0.044 & 30 &  $\mathtt{hd{\mbox{-}}11{\mbox{-}}d0{\mbox{-}}prog}$  &  0.172   & 0.016 \\
      03 &  $\mathtt{hd{\mbox{-}}02{\mbox{-}}d0{\mbox{-}}prog}$  &  0.070   & 0.017 & 31 &  $\mathtt{hd{\mbox{-}}11{\mbox{-}}d1{\mbox{-}}prog}$  &  0.742   & 0.019 \\
      04 &  $\mathtt{hd{\mbox{-}}02{\mbox{-}}d1{\mbox{-}}prog}$  &  0.123   & 0.018 & 32 &  $\mathtt{hd{\mbox{-}}11{\mbox{-}}d5{\mbox{-}}prog}$  &  3.402   & 6.318 \\
      05 &  $\mathtt{hd{\mbox{-}}02{\mbox{-}}d5{\mbox{-}}prog}$  &  0.175   & 0.030 & 33 &  $\mathtt{hd{\mbox{-}}12{\mbox{-}}d0{\mbox{-}}prog}$  &  0.371   & 0.014 \\
      06 &  $\mathtt{hd{\mbox{-}}03{\mbox{-}}d0{\mbox{-}}prog}$  &  0.033   & 0.013 & 34 &  $\mathtt{hd{\mbox{-}}12{\mbox{-}}d1{\mbox{-}}prog}$  &  0.418   & 0.018 \\
      07 &  $\mathtt{hd{\mbox{-}}03{\mbox{-}}d1{\mbox{-}}prog}$  &  0.060   & 0.014 & 35 &  $\mathtt{hd{\mbox{-}}12{\mbox{-}}d5{\mbox{-}}prog}$  &  0.969   & 0.025 \\
      08 &  $\mathtt{hd{\mbox{-}}03{\mbox{-}}d5{\mbox{-}}prog}$  &  0.089   & 0.014 & 36 &  $\mathtt{hd{\mbox{-}}13{\mbox{-}}d0{\mbox{-}}prog}$  &  0.341   & 0.036 \\
      09 &  $\mathtt{hd{\mbox{-}}04{\mbox{-}}d0{\mbox{-}}prog}$  &  0.039   & 0.014 & 37 &  $\mathtt{hd{\mbox{-}}13{\mbox{-}}d1{\mbox{-}}prog}$  &  1.825   & 0.775 \\
      10 &  $\mathtt{hd{\mbox{-}}04{\mbox{-}}d1{\mbox{-}}prog}$  &  0.062   & 0.025 & 38 &  $\mathtt{hd{\mbox{-}}13{\mbox{-}}d5{\mbox{-}}prog}$  &  5.418   & 8.305 \\
      11 &  $\mathtt{hd{\mbox{-}}04{\mbox{-}}d5{\mbox{-}}prog}$  &  0.078   & 0.048 & 39 &  $\mathtt{hd{\mbox{-}}14{\mbox{-}}d0{\mbox{-}}prog}$  &  0.523   & 0.091 \\
      12 &  $\mathtt{hd{\mbox{-}}05{\mbox{-}}d0{\mbox{-}}prog}$  &  0.040   & 0.022 & 40 &  $\mathtt{hd{\mbox{-}}14{\mbox{-}}d1{\mbox{-}}prog}$  &  9.770   & 8.151 \\
      13 &  $\mathtt{hd{\mbox{-}}05{\mbox{-}}d1{\mbox{-}}prog}$  &  0.127   & 0.022 & 41 &  $\mathtt{hd{\mbox{-}}14{\mbox{-}}d5{\mbox{-}}prog}$  &  TO      & TO    \\
      14 &  $\mathtt{hd{\mbox{-}}05{\mbox{-}}d5{\mbox{-}}prog}$  &  0.155   & 0.044 & 42 &  $\mathtt{hd{\mbox{-}}15{\mbox{-}}d0{\mbox{-}}prog}$  &  0.561   & 0.164 \\
      15 &  $\mathtt{hd{\mbox{-}}06{\mbox{-}}d0{\mbox{-}}prog}$  &  0.034   & 0.020 & 43 &  $\mathtt{hd{\mbox{-}}15{\mbox{-}}d1{\mbox{-}}prog}$  &  3.586   & 6.164 \\
      16 &  $\mathtt{hd{\mbox{-}}06{\mbox{-}}d1{\mbox{-}}prog}$  &  0.091   & 0.020 & 44 &  $\mathtt{hd{\mbox{-}}15{\mbox{-}}d5{\mbox{-}}prog}$  &  TO      & TO    \\
      17 &  $\mathtt{hd{\mbox{-}}06{\mbox{-}}d5{\mbox{-}}prog}$  &  0.142   & 0.033 & 45 &  $\mathtt{hd{\mbox{-}}17{\mbox{-}}d0{\mbox{-}}prog}$  &  0.788   & 0.085 \\
      18 &  $\mathtt{hd{\mbox{-}}07{\mbox{-}}d0{\mbox{-}}prog}$  &  0.103   & 0.017 & 46 &  $\mathtt{hd{\mbox{-}}17{\mbox{-}}d1{\mbox{-}}prog}$  &  1.210   & 0.119 \\
      19 &  $\mathtt{hd{\mbox{-}}07{\mbox{-}}d1{\mbox{-}}prog}$  &  0.238   & 0.027 & 47 &  $\mathtt{hd{\mbox{-}}17{\mbox{-}}d5{\mbox{-}}prog}$  &  3.184   & 4.725 \\
      20 &  $\mathtt{hd{\mbox{-}}07{\mbox{-}}d5{\mbox{-}}prog}$  &  0.281   & 0.044 & 48 &  $\mathtt{hd{\mbox{-}}18{\mbox{-}}d0{\mbox{-}}prog}$  &  0.121   & 0.019 \\
      21 &  $\mathtt{hd{\mbox{-}}08{\mbox{-}}d0{\mbox{-}}prog}$  &  0.077   & 0.018 & 49 &  $\mathtt{hd{\mbox{-}}18{\mbox{-}}d1{\mbox{-}}prog}$  &  0.382   & 0.157 \\
      22 &  $\mathtt{hd{\mbox{-}}08{\mbox{-}}d1{\mbox{-}}prog}$  &  0.223   & 0.029 & 50 &  $\mathtt{hd{\mbox{-}}18{\mbox{-}}d5{\mbox{-}}prog}$  &  0.586   & 0.044 \\
      23 &  $\mathtt{hd{\mbox{-}}08{\mbox{-}}d5{\mbox{-}}prog}$  &  0.281   & 0.027 & 51 &  $\mathtt{hd{\mbox{-}}19{\mbox{-}}d0{\mbox{-}}prog}$  &  TO      & TO    \\
      24 &  $\mathtt{hd{\mbox{-}}09{\mbox{-}}d0{\mbox{-}}prog}$  &  0.385   & 0.014 & 52 &  $\mathtt{hd{\mbox{-}}19{\mbox{-}}d1{\mbox{-}}prog}$  &  TO      & TO    \\
      25 &  $\mathtt{hd{\mbox{-}}09{\mbox{-}}d1{\mbox{-}}prog}$  &  0.972   & 0.418 & 53 &  $\mathtt{hd{\mbox{-}}19{\mbox{-}}d5{\mbox{-}}prog}$  &  TO      & TO    \\
      26 &  $\mathtt{hd{\mbox{-}}09{\mbox{-}}d5{\mbox{-}}prog}$  &  1.485   & 0.573 & 54 &  $\mathtt{hd{\mbox{-}}20{\mbox{-}}d0{\mbox{-}}prog}$  &  TO      & TO    \\
      27 &  $\mathtt{hd{\mbox{-}}10{\mbox{-}}d0{\mbox{-}}prog}$  &  0.142   & 0.012 & 55 &  $\mathtt{hd{\mbox{-}}20{\mbox{-}}d1{\mbox{-}}prog}$  &  TO      & TO    \\
      28 &  $\mathtt{hd{\mbox{-}}10{\mbox{-}}d1{\mbox{-}}prog}$  &  0.542   & 0.021 & 56 &  $\mathtt{hd{\mbox{-}}20{\mbox{-}}d5{\mbox{-}}prog}$  &  TO      & TO    \\
      \hline
    \end{tabular}
    \caption{Results for Section~\ref{sec:bv}\label{table:bv_results}}
  \end{table}

We implemented the above procedure in a tool called {\PUFFIN} 
and evaluated it on benchmarks from the linear
integer arithmetic track with separable specifications from the SyGuS
competition 2014.
The $\mathtt{max}_n$ benchmarks specify a function that outputs the
maximum of $n$ input variables (the illustrative example from
Section~\ref{sec:illustrative_example} is $\max_2$). Note that
the SyGuS competition benchmarks only go up to $\max_5$.
The $\mathtt{array\_search}_n$ and $\mathtt{array\_sum}_n$ benchmarks
respectively specify functions that search for a given input in a sorted
array, and check if the sum of two consecutive elements in an array 
is equal to a given value.
In all these benchmarks, our tool significantly outperforms {\esolver}
and other the existing solvers based on a pure CEGIS approach.
The reason for this is as follows: the CEGIS solvers try to generate
the whole program at once, which is a complex expression. 
On the other hand, our solver combines simple expressions generated for
parts of the input spaces where the output expression is simple.

  \newpage
  \section{Appendix to Section~\ref{sec:lia_nonfunc}}
    \begin{table}
  \centering
      \begin{tabular}{|c|c|r|r|}
        \hline Sl. no & Benchmark        & {\RAZORBILL}     & {\esolver} \\
                      &                  & {Time (sec)}  &  Time (sec)    \\
        \hline
        \hline
        01 &  $\mathtt{unbdd\_inv\_gen\_array}$  &   2.855   &  0.046 \\
        02 &  $\mathtt{unbdd\_inv\_gen\_cegar2}$ &   0.742   &  10.579 \\
        03 &  $\mathtt{unbdd\_inv\_gen\_cgr1}$   &   0.659   &  0.022 \\
        04 &  $\mathtt{unbdd\_inv\_gen\_ex14}$   &   0.490   &  0.013 \\
        05 &  $\mathtt{unbdd\_inv\_gen\_ex23}$   &   0.369   & 70.105 \\
        06 &  $\mathtt{unbdd\_inv\_gen\_ex7}$    &   0.517   &  0.026 \\
        07 &  $\mathtt{unbdd\_inv\_gen\_fig1}$   &   2.341   &  0.148 \\
        08 &  $\mathtt{unbdd\_inv\_gen\_fig3}$   &   2.910   &  0.128 \\
        09 &  $\mathtt{unbdd\_inv\_gen\_fig6}$   &   0.812   &  0.009 \\
        10 &  $\mathtt{unbdd\_inv\_gen\_fig8}$   &   0.519   &  0.011 \\
        11 &  $\mathtt{unbdd\_inv\_gen\_fig9}$   &   0.298   &  0.014 \\
        12 &  $\mathtt{unbdd\_inv\_gen\_finf1}$  &   0.700   &  0.010 \\
        13 &  $\mathtt{unbdd\_inv\_gen\_finf2}$  &   0.620   &  0.011 \\
        14 &  $\mathtt{unbdd\_inv\_gen\_n\_c11}$ &   0.358   &  0.022 \\
        15 &  $\mathtt{unbdd\_inv\_gen\_sum1}$   &   0.444   &  23.555 \\
        16 &  $\mathtt{unbdd\_inv\_gen\_sum3}$   &   0.088   &  0.014 \\
        17 &  $\mathtt{unbdd\_inv\_gen\_sum4}$   &   0.959   &  5.478 \\
        18 &  $\mathtt{unbdd\_inv\_gen\_tcs}$    &   0.835   &  26.810 \\
        19 &  $\mathtt{unbdd\_inv\_gen\_term2}$  &   0.602   &  0.015 \\
        20 &  $\mathtt{unbdd\_inv\_gen\_term3}$  &   0.887   &  0.013 \\
        21 &  $\mathtt{unbdd\_inv\_gen\_trex1}$  &   0.305   &  0.017 \\
        22 &  $\mathtt{unbdd\_inv\_gen\_trex2}$  &   0.014   &  0.011 \\
        23 &  $\mathtt{unbdd\_inv\_gen\_trex4}$  &   0.298   &  0.011 \\
        24 &  $\mathtt{unbdd\_inv\_gen\_vmail}$  &   0.925   &  0.012 \\
        25 &  $\mathtt{unbdd\_inv\_gen\_w1}$     &   0.725   &  0.012 \\
        26 &  $\mathtt{unbdd\_inv\_gen\_w2}$     &   0.471   &  0.027 \\
        27 &  $\mathtt{unbdd\_inv\_gen\_winf1}$  &   0.707   &  0.010 \\
        28 &  $\mathtt{unbdd\_inv\_gen\_winf2}$  &   0.578   &  0.015 \\
        \hline
      \end{tabular}
      \caption{Results for Section~\ref{sec:lia_nonfunc}.
      \label{table:lia_nr_results}}
    \end{table}

We implemented the above procedure in a tool called {\RAZORBILL}
and evaluated it linear integer benchmarks with non-separable
specifications from SyGuS competition 2014.
We compare the performance of our tool and {\esolver} on the $29$ invgen
set of benchmarks (results in Table~\ref{table:lia_nr_results} in the
appendix).
As in the case of bit-vector benchmarks, on small and easy benchmarks
(where both tools return the solution in less than $1$ second),
{\esolver} is faster.
However, on larger benchmarks, {\RAZORBILL} can be much faster (see for
example, benchmark {\tt inv\_gen\_ex23}).
As before, we hypothesize that this is due to the enumerative {\esolver}
quickly enumerating small solutions before the STUN based solver can
perform any complex theory reasoning.

\end{appendix}

\fi

\end{document}